# HCI and Educational Metrics as Tools for VLE Evaluation

Vita Hinze-Hoare
School of Electronics and Computer Science
University of Southampton
United Kingdom

**Abstract**
*The general set of HCI and Educational principles are considered and a classification system constructed. A frequency analysis of principles is used to obtain the most significant set. Metrics are devised to provide objective measures of these principles and a consistent testing regime devised. These principles are used to analyse Blackboard and Moodle.*

## 1. Introduction

A well-known problem[1] with the evaluation of educational virtual learning environments (VLEs) is the lack of a clear objective assessment framework having a wide recognition. This means that there is an issue over the best way of evaluating their effectiveness on both sound educational principles and on Human Computer Interface principles.

It is the aim of this paper to highlight some of the steps to move toward an objective standard by which to gauge VLEs and ultimately to provide a single overall index measure (essentially a score out of 10) for both usability and educational worth based upon an analysis of accepted standards. An HCI index was constructed for general usability comparison and a separate educational index (EDI index) was designed to provide a measure of educational quality.

First the Blackboard VLE and second an open source VLE, Moodle, were tested. As far as possible the open source VLE carried the same content as the Blackboard VLE to allow a comparison of the VLE structure and operation rather than its content. Usability statistics are obtained from a set of standard users.

## 2. Possible Frameworks and Approaches

Appropriate HCI and Educational Frameworks were constructed to provide a measure of the VLE in these areas. There are a number of competing HCI principles championed by various experts in that field, but no clear evaluation framework exists as yet.

An analysis of the literature produced by the leading authors in the field lead to the listing of the HCI principles most commonly promoted by performing a frequency analysis of principles and then of authors from which a definitive list was established.

In the area of educational principles the approach is not so straightforward as there are many more competing pedagogical frameworks from a vast array of authors. Sandy Britain 1998 working on behalf of JISC has evaluated VLEs on the basis of the conversational framework developed by Laurillard 1993[2]. This provided a medium for supporting constructivist and conversational approaches to learning with the central emphasis being placed upon dialogue between student and teacher. The weakness of this work is in the limited acceptance of Laurillard's methodology.

At the outset an analysis was performed on educational principles following a similar method of that adopted for the HCI principles. It was discovered that whereas the principles of the HCI were mutually inclusive and could therefore be adopted in conjunction with each other, the Educational principles from different educational theorists were mutually exclusive and could not be mixed and matched in the same way. Each Educational principle was constructed as part of the edifice of a single educational theory and could not be removed and placed with a different principle in a different theory. A new approach was necessary and it was decided to adopt instead the most prevailing educational theory today from a single theoretical school.

An approach was considered that centred upon the adoption of a particular framework from one of the most highly cited educational theorist of the day namely Jerome Bruner[3] supported by and based upon the work of Vygotsky was considered.

## 3. HCI Principles

A set of 8 principles that could be confidently represented as the most significant and widely supported in the field was determined.

The following table shows the relative frequency of citation by the leading authors in order of significance. Seventeen different sources were consulted.[4]

| 1 | Familiarity | 12 |
|---|---|---|
| 2 | Consistency | 12 |
| 3 | Forward Error Recovery | 9 |
| 4 | Substitutivity | 8 |
| 5 | Dialogue Initiative | 7 |
| 6 | Task Migratablility | 7 |
| 7 | Responsiveness | 6 |
| 8 | Customisability | 6 |

**4. Educational Principles**

By adopting Bruner's approach the education principles are obtained from one of his seminal works on educational theory, "Process of Education" in which he presents the essential components for effective learning.

Bruner's principles are listed below and these are widely adopted by the constructivist school of thought which can be argued is the predominant educational theory today. The five principles listed in the table below were then used to construct appropriate questions in a user survey to enable a measure of the EDI index.

|  | **Tools** | **Structuring** |
|---|---|---|
| **Collaboration** | What tools assist? Forums, Chat, email, shared work areas? | Can discussions easily be put together? |
| **Control** | Can the student tailor the system in a way which is personally beneficial | Is the structure amenable to flexible approaches? |
| **Culture** | How does the VLE engender a culture of learning? What tools are used for this? | Does the structure of the VLE foster a sense of learning identity? |
| **Reflection** | What tools promote reflection? Journals, Mind mapping? | How does the VLE promote meta-cognition? Learning about learning? |
| **Reinforcement** | What tools reinforce learning? | How does the VLE establish the learning in the mind of the student? |

**5. Overview of Moodle and Blackboard**

|  | **Moodle** | **Blackboard** |
|---|---|---|
| **Structure** | Three formats, weekly, topical and social | Hierarchical and sequential structuring |
| **Learning Activities** | Connected discussions, Evaluation and rating of comments, Reflective journals, glossary, Encyclopaedia, writing, Chatting, peer-evaluated assignments | Discussion boards, chat rooms, assessments, grade books, announcement editors, virtual institutions |
| **Time Management and Planning** | Calendar, Activity Report | Calendar, Tasks and Notes, Portfolios |
| **Pedagogical Approach** | Constructivist philosophy | None |
| **Critique** | Haphazard and unplanned development, avoid exposure to vendor lock-in, Moodle does not offer any program view of courses nor does it currently afford students the capability of organising themselves or maintaining a presence outside the VLE | Does not connect things in an instructionally useful way. "Learning environment is trivia orientated" Tracking facilities doe not provide useful information. |

**6. Set up of Open source VLE**

The online Moodle VLE involved the setting up of a MYSQL database to store all VLE data. Installation involved running a number of PHP scripts, which installed the relevant files in the correct directories and attached connections to the MSQL database. A configuration file was set up to point to the appropriate location, and index pages were created to provide the appropriate link.

**7. Evaluation of both VLEs**

A Survey questionnaire was created. There were 34 questions in all, which were randomized between HCI and Educational Principles.
About 80 people from Southampton University were invited to partake. There were 8 participants in total. The survey took 6 minutes on average to complete and was conducted on-line and an initial analysis provided by in-built tools in Moodle. Further analysis was performed with Excel.

**8. Analysis of Results**

The Survey was embedded into Moodle, which automatically listed the name of each participant together with their overall percentage grade comparing the two VLEs This was constructed in such a way that percentages over 50% were more in favour of the Moodle VLE up to a maximum of 100% and conversely those below 50% in favour of Blackboard down to a value of 0%. This crude overall grade gives a very simple macro view of the relative perceived merits of each VLE.

The second analysis provided the number of responses for each individual answer to the 34 survey questions. This item response analysis allows us to determine the relative strengths and weaknesses of each VLE in any of the five Educational and eight HCI areas.

The third analysis performed was a listing of items in the survey with summary statistics. This provided the percentage breakdown of answers for each question. This enabled a relative determination of the strength of each VLE down to the individual question level.

The automatic facilities further provided the downloading of those statistics in excel format for further analysis. This included the accumulative number of answers for the total number of question which provides a statistical distribution showing the balance of choices in favour of one or the other VLE in both the Educational and HCI areas.

**Specific Educational Results**
The initial result showed that the Blackboard VLE scored 5.1, which was fractionally more the Moodle VLE at 4.9 in terms of educational values. This is a difference by a factor of 2% and suggests that within the accuracy of the study Moodle and Blackboard equally meets the educational needs of the students.

Blackboard scored more highly in the areas of learner control (+2.25%) and reflective learning (+0.75%). Although not as significant as the HCI differences they nevertheless indicate that they had more control over the learning process with Blackboard than with Moodle and that Blackboard enables them to think more about the process of learning.
The only area where Moodle scored more highly was in the cultural environment of learning (+0.9%), which suggests that Moodle fosters a holistic cultural identity in a slightly more effective way than Blackboard.

The frequency analysis for the Educational index indicates slight skewing in favour of Moodle as opposed to Blackboard.

**Summary Specific Educational results**

|  | Moodle | Rel. score |
|---|---|---|
| Collaborative Learning | 49.76% | -0.24% |
| Learner Control | 47.78% | -2.22% |
| Reflective Learning | 49.24% | -0.76% |
| Cultural Learning | 50.95% | 0.95% |
| Reinforcement | 49.71% | -0.29% |
| **TOTAL ED REL INDEX** | **49.49%** | **-0.51%** |

Moodle fared best with
Cultural learning +0.9%

Blackboard fared best with
Learner control + 2.25%
Reflective learning + 0.75%
Reinforcement + 0.25%
Collaborative learning + 0.25%

**Specific HCI Results**
The second result showed that the Moodle VLE scored 4.7 as against Blackboard's score of 5.3 in the area of HCI implementation. This is a difference of 0.6% and suggests that within the accuracy of the study Blackboard and Moodle equally meet the usability needs of the students.

Blackboard scored most highly in the areas of task migratability (+ 16%) and responsiveness (+9%), whereas Moodle most strongly featured dialogue initiative (+3%). This suggests that Blackboard is perceived as the most efficient system for providing a high rate of communication between the user and the system. In addition to this, the high task migratability score suggests that users find that they have more control over the system. On the other hand the slightly elevated score for Moodle in dialogue initiative suggests that users felt that the dialogue interface was more effective and helpful.

The frequency analysis see results table 4 for HCI gives a slightly more interesting picture indicating a slight skewing towards the Moodle end when the weighting factors are not taken into account. This shows that apportioning of percentages to the various answers might have an effect upon the overall result. The simple frequency analysis of the numbers of questions answered shows a slight favour in terms of Moodle.

**Summary Specific individual HCI results**

|  | Moodle | Rel. Score |
|---|---|---|
| Familiarity | 50.68% | 0.68% |
| Consistency | 50.51% | 0.51% |

Page 3 of 10

| | | |
|---|---|---|
| Customizability | 48.28% | -1.72% |
| Responsiveness | 40.60% | -9.40% |
| Dialogue In | 53.10% | 3.10% |
| Task Migratability | 34.40% | -15.60% |
| Error Recovery | 51.29% | 1.29% |
| Substitutivity | 50.80% | 0.80% |
| **TOTAL HCI REL INDEX** | **47.46%** | **-2.54%** |

Moodle fared best with
Dialogue initiative +2.5%
Error recovery +1%
Substitutivity +.0.3%
Familiarity +.0.3%
Consistency +0.25%

Blackboard fared best with
Task Migratability +16%
Responsiveness +9%
Customisability +2%^

**Conclusions from the Survey**

Both VLEs are perceived by this survey to contain weaknesses in certain areas, which could benefit from further consideration by the creators. This includes attention to consistency in use and recoverability from mistakes and feedback information via dialogue boxes and ease of understanding and providing more choice in the way to accomplish various tasks. These are issues, which are primarily HCI considerations, and the suggestion is that Moodle is slightly weaker in its implementation of HCI principles.

Moodle is perceived by this survey to contain weaknesses in the areas of both familiarity of use and expected responses. This is not an unsurprising result as all of the users in the survey had continuous and persistent use of the Blackboard VLE over a number of years whereas their introduction to the Moodle VLE was in most cases expected to be brief and new.
It would therefore be expected that in the area of familiarity Blackboard would score more highly with this sample of users. As such this result cannot be taken as establishing anything significant. In order to provide a significant result it would be required to find groups of users who had no previous experience of either VLE in question.

Moodle was rated the better VLE as far as cultural learning was concerned, providing better community resources.
Blackboard was thought better for Task Migratability and Responsiveness
The surprising result is that Moodle scored fractionally higher than Blackboard in the area of familiarity, which is an interesting result considering that Blackboard has been used by all participants for a much longer period.

**Conclusions for future development**

The survey could be extended to examine the back office workings as well as the front shop by including a survey of course designers and lecturers.

Additional testing might be done on a population of users, which had no prior familiarity with either VLE for the purposes of ensuring a lack of bias.

**APPENDIX: Graphical Results**

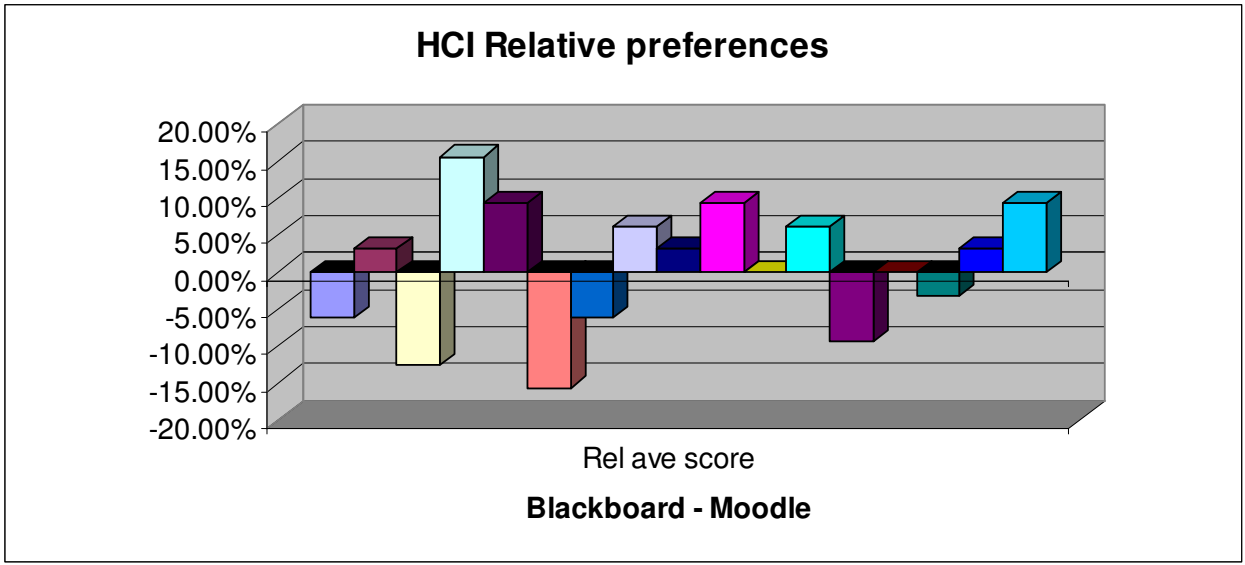

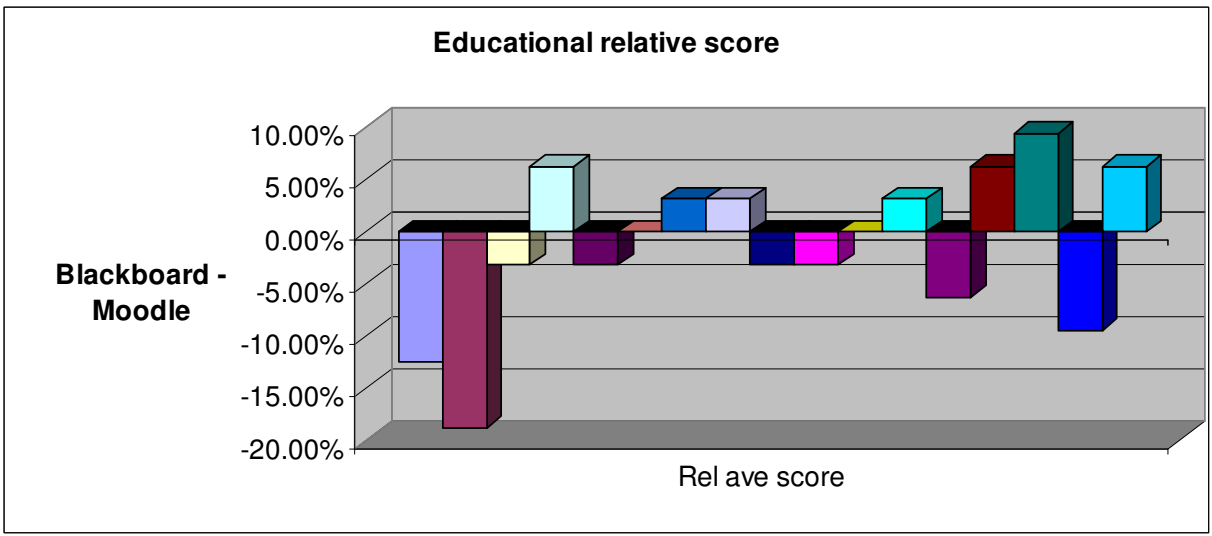

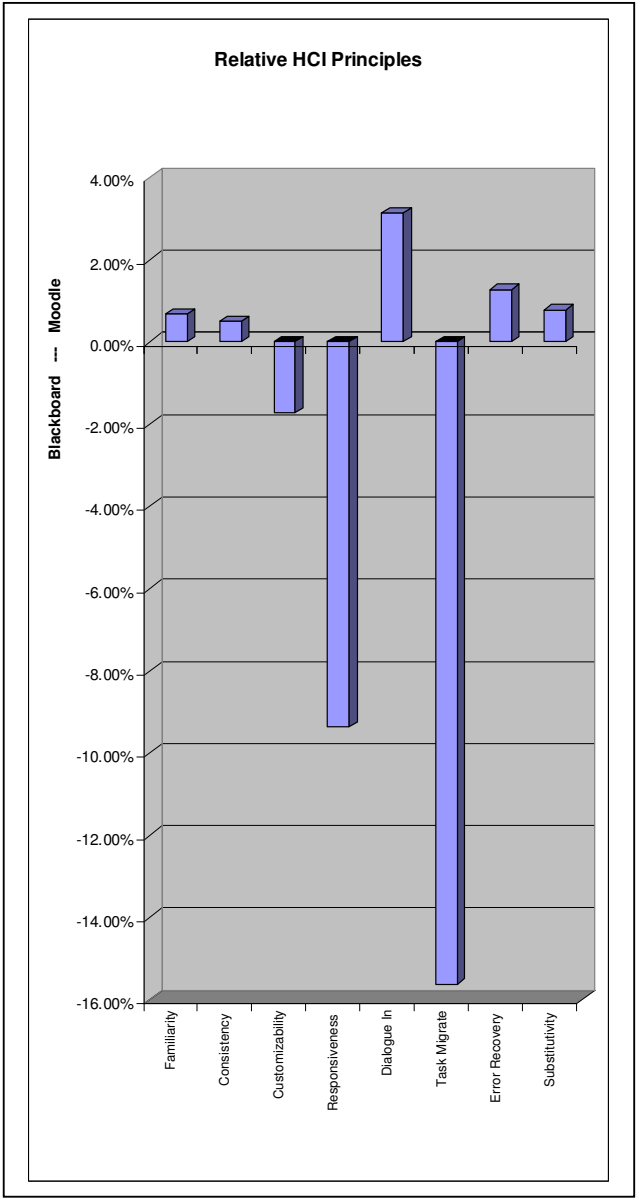
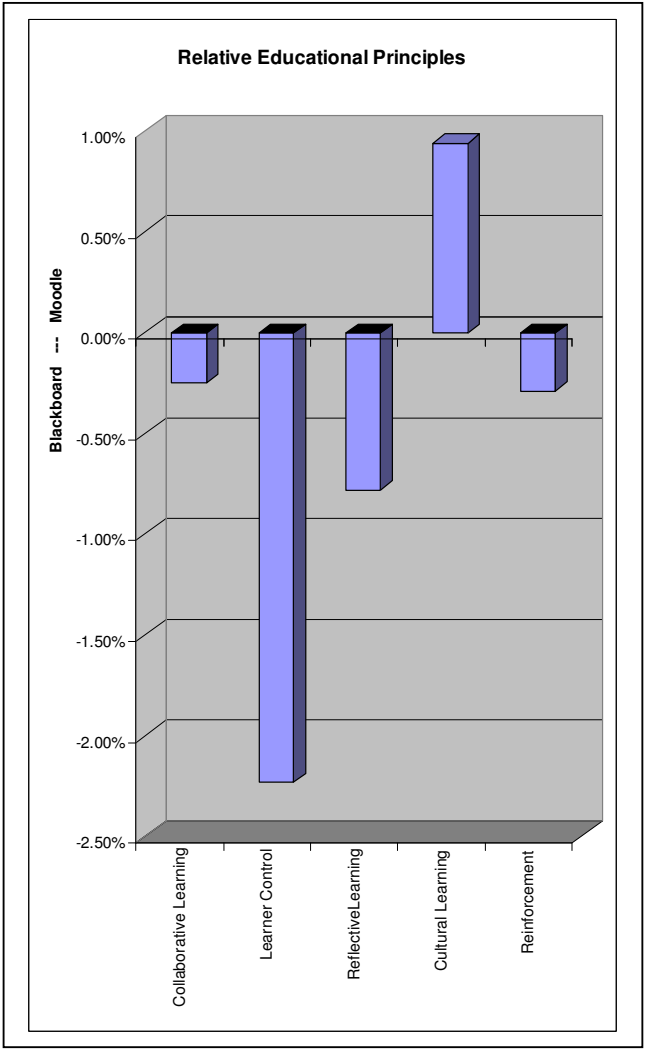

### Results Table 4

| Question Number | HCI | HCI Type | Ed Type | Ave Score | Rel ave score |
|---|---|---|---|---|---|
| 1 | HCI | Customizability | | 43.80% | -6.20% |
| 2 | Educational | | Learner Control | 37.50% | -12.50% |
| 3 | Educational | | Collaborative Learning | 31.30% | -18.70% |
| 4 | HCI | Dialogue initiative | | 53.10% | 3.10% |
| 5 | Educational | | Cultural Learning | 46.90% | -3.10% |
| 6 | HCI | Familiarity | | 37.50% | -12.50% |
| 7 | Educational | | Learner Control | 56.30% | 6.30% |
| 8 | HCI | Error Recovery | | 65.60% | 15.60% |
| 9 | HCI | Familiarity | | 59.40% | 9.40% |
| 10 | Educational | | Reinforcement | 46.90% | -3.10% |
| 11 | HCI | Task Migratability | | 34.40% | -15.60% |
| 12 | Educational | | Reinforcement | 50.00% | 0.00% |
| 13 | HCI | Consistency | | 43.80% | -6.20% |
| 14 | Educational | | Learner Control | 53.10% | 3.10% |
| 15 | Educational | | Learner Control | 53.10% | 3.10% |
| 16 | Educational | | Collaborative Learning | 46.90% | -3.10% |
| 17 | HCI | Consistency | | 56.30% | 6.30% |
| 18 | HCI | Customizability | | 53.10% | 3.10% |
| 19 | HCI | Consistency | | 59.40% | 9.40% |
| 20 | HCI | Substitutivity | | 50.00% | 0.00% |
| 21 | Educational | | Cultural Learning | 46.90% | -3.10% |
| 22 | Educational | | Collaborative Learning | 50.00% | 0.00% |
| 23 | HCI | Substitutivity | | 56.30% | 6.30% |
| 24 | Educational | | Cultural Learning | 53.10% | 3.10% |
| 25 | Educational | | ReflectiveLearning | 43.80% | -6.20% |
| 26 | HCI | Responsiveness | | 40.60% | -9.40% |
| 27 | Educational | | Collaborative Learning | 56.30% | 6.30% |
| 28 | Educational | | Reinforcement | 59.40% | 9.40% |
| 29 | HCI | Familiarity | | 50.00% | 0.00% |
| 30 | Educational | | Reinforcement | 40.60% | -9.40% |
| 31 | HCI | Consistency | | 46.90% | -3.10% |
| 32 | Educational | | ReflectiveLearning | 56.30% | 6.30% |
| 33 | HCI | Familiarity | | 53.10% | 3.10% |
| 34 | HCI | Error Recovery | | 59.40% | 9.40% |

| | | |
|---|---|---|
| Familiarity | 50.68% | 0.68% |
| Consistency | 50.51% | 0.51% |
| Customizability | 48.28% | -1.72% |
| Responsiveness | 40.60% | -9.40% |
| Dialogue In | 53.10% | 3.10% |
| Task Migrate | 34.40% | -15.60% |
| Error Recovery | 51.29% | 1.29% |
| Substitutivity | 50.80% | 0.80% |
| **TOTAL HCI REL INDEX** | **47.46%** | **-2.54%** |
| Collaborative Learning | 49.76% | -0.24% |
| Learner Control | 47.78% | -2.22% |
| Reflective Learning | 49.24% | -0.76% |
| Cultural Learning | 50.95% | 0.95% |
| Reinforcement | 49.71% | -0.29% |
| **TOTAL ED REL INDEX** | **49.49%** | **-0.51%** |

| Moodle | 4.7 | Blackboard | 5.3 | Moodle | 4.9 | Blackboard | 5.1 |

End Notes

[1] Nielsen, J., (2000) *Designing Web Usability*, New Riders Publishing, P13
[2] Laurillard, D. (1993) *Rethinking University Teaching – a framework for the effective use of educational technology,* London: Routledge.
[3] Bruner, J., *http://tip.psychology.org/bruner.html*
[4] Apendix 1, Reflective Report.